\documentclass[preprint,12pt]{aastex}

\newcommand{\chandra}{{\it Chandra}}
\newcommand{\xmm}{{\it XMM}}
\newcommand{\asca}{{\it ASCA}}
\newcommand{\ginga}{{\it Ginga}}
\newcommand{\ee}[1]{\times10^{#1}}
\newcommand{\lya}{Ly$\alpha$}
\newcommand{\lyaone}{Ly$\alpha_1$}
\newcommand{\lyatwo}{Ly$\alpha_2$}
\newcommand{\mdot}{\dot{M}}
\newcommand{\vinf}{v_\infty}
\newcommand{\kms}{km\,s$^{-1}$}
\newcommand{\mump}{\mu m_p}
\newcommand{\ergs}{erg\,s$^{-1}$}

\newcommand{\ud}[2]{$^{+#1}_{-#2}$}
\newcommand{\ig}[2]{\includegraphics[angle=-90,width=#1in]{#2}}
\newcommand{\glnoscat}[1]{}

\newcommand{\nuth}{{\nu_{\rm th}}}
\newcommand{\nulll}[1]{}
\shorttitle{Resonant Scattering in Cen~X-3}
\shortauthors{Wojdowski, et al.}
\begin{document}

\title{Resolving the Effects of Resonant X-ray Line Scattering in
Cen~X-3 with {\it Chandra}}
\author{Patrick S. Wojdowski\altaffilmark{1}, Duane A. Liedahl}
\affil{Department of Physics and Advanced Technologies,
Lawrence Livermore National Laboratory}
\affil{P.O. Box 808, Livermore, CA 94551}
\email{pswoj@space.mit.edu,duane@virgo.llnl.gov}
\and
\author{Masao Sako\altaffilmark{2}, Steven M. Kahn, Frederik Paerels}
\affil{Columbia Astrophysics Laboratory and Department of Physics,
Columbia University, 538 West 120th St., New York, NY 10027}
\email{masao@tapir.caltech.edu, skahn@astro.columbia.edu, 
frits@astro.columbia.edu}
\altaffiltext{1}{Current Address: Center for Space Research,
Massachusetts Institute of Technology, NE80-6003, 77 Massachusetts
Ave., Cambridge, MA 02139}
\altaffiltext{2}{Chandra Fellow, Current Address: Theoretical
Astrophysics and Space Radiation Laboratory, California Institute of
Technology, MC 130-33, Pasadena, CA 91125}

\begin{abstract}
The massive X-ray binary Cen~X-3 was observed over approximately one
quarter of the system's 2.08 day orbit, beginning before eclipse and
ending slightly after eclipse center with the {\it Chandra X-ray
Observatory} using its High-Energy Transmission Grating Spectrometer.
The spectra show K shell emission lines from hydrogen- and helium-like
ions of magnesium, silicon, sulfur, and iron as well as a K$\alpha$
fluorescence emission feature from near-neutral iron.  The helium-like
$n=2\to1$ triplet of silicon is fully resolved and the analogous
triplet of iron is partially resolved.  We measure fluxes, shifts, and
widths of the observed emission lines.  The helium-like triplet
component flux ratios outside of eclipse are consistent with emission
from recombination and subsequent cascades (recombination radiation)
from a photoionized plasma with temperature $\sim$100\,eV.  In
eclipse, however, the $w$ (resonance) lines of silicon and iron are
stronger than that expected for recombination radiation, and are
consistent with emission from a collisionally ionized plasma with a
temperature of $\sim$1\,keV.  The triplet line flux ratios at both
phases can be explained more naturally, however, as emission from a
photoionized plasma if the effects of resonant line scattering, a
process which has generally been neglected in X-ray spectroscopy, are
included in addition to recombination radiation.

We show that resonant line scattering in photoionized plasmas may
increase the emissivity of $n=2\to1$ line emission in hydrogen and
helium-like ions by a factor as large as four relative to that of pure
recombination and so previous studies, in which resonant scattering
has been neglected, may contain significant errors in the derived
plasma parameters.  The emissivity due to resonance scattering depends
sensitively on the line optical depth and, in the case of winds in
X-ray binaries, this allows constraints on the wind velocity even
when Doppler shifts cannot be resolved.

\end{abstract}

\section{Introduction}
\label{sec:intro}

It has long been known that in eclipsing high mass X-ray binaries
(HMXBs, X-ray binaries where the companion star is a massive star of
type O or B) a residual X-ray flux can be observed during eclipse of
the compact X-ray source \citep{sch72}.  \citet{bec78} observed that
in Vela~X-1, the spectral shape of the hard X-ray flux observed during
eclipse was approximately the same as that observed out of eclipse,
and inferred from this that the residual eclipse flux was due to
electron scattering of the primary X-rays in circumstellar material.
The residual eclipse fluxes of X-ray binaries are typically a few
percent of the out-of-eclipse fluxes, implying that the scattering
optical depth in the continuum is approximately a few percent.  Using
this electron scattering optical depth and, for the length scale, the
orbital separation $a$ (of order $10^{12}$\,cm) circumstellar
densities of order $10^{10}$\,cm$^{-3}$ are inferred, assuming
$\tau=n_e\sigma_{\rm T}a$, where $\sigma_{\rm T}$ is the Thomson
cross-section.

Isolated O and B type stars have winds with mass-loss rates of order
$10^{-7}$--$10^{-6}\,M_\sun$\,yr$^{-1}$ and velocities of order
1000\,\kms.  The 
winds are driven by ultraviolet photons which impart their outward
momentum to the matter primarily through scattering in line
transitions \citep{luc70,cas75}.  For a spherically symmetric steady-state
wind, the hydrogen atom density ($n$) is related to the mass-loss rate
($\mdot$) and velocity ($v$) by
\begin{equation}
n=\frac{\mdot}{4\pi\mump r^2v}
\label{eqn:n}
\end{equation}
where $r$ is the distance from the stellar center, $m_p$ is the proton
mass, and we take $\mu$, the mass in amu per hydrogen atom, to be 1.4.
Using, again, the length scale $10^{12}$\,cm, the average density
derived above from electron scattering is consistent with the
densities in the winds of massive stars.

This similarity suggests that the dynamics of hot star winds in X-ray
binary systems might be quite similar to the those in isolated stars.
However, the X-ray luminosities ($L$) of these objects are of order
$10^{37}$\,\ergs, implying that typical values of the ionization
parameter \citep[$\xi=L/ns^2$, where $s$ is the distance from the
X-ray source,][]{tar69} in the circumstellar material are of order
$10^3$\,\ergs{}cm, meaning that the wind should be highly ionized.  At
such high ionization, the UV line opacity of the wind is greatly
reduced and, therefore, so is the force exerted on the wind by the
star's radiation \citep{ste90}.  Several calculations (which, however,
have several significant approximations) have shown that UV photons
cannot drive a wind on the X-ray illuminated side of an X-ray luminous
HMXB \citep[e.g.,][]{ste91}.  It may be that an alternative mechanism,
such as evaporation from the X-ray heated surface of the companion
\citep[e.g.,][]{bas77} drives the wind in luminous HMXBs.

The high ionization conditions in the winds of HMXBs can be inferred
directly by the observation of K shell emission lines from hydrogen-
and helium-like ions.  \citet{nag92}, for example, noticed in a
\ginga\ observation of Cen~X-3 that the iron line was
not consistent with only a 6.4\,keV line from neutral iron but could
be fit if a 6.7\,keV component from helium-like iron was included.
The launch of the \asca\ observatory, with its Solid-State Imaging
Spectrometers (SIS), represented a great improvement over previous
observatories in the sensitivity to X-ray emission lines.  With the SIS
instrument, winds of HMXBs were seen to produce K-shell emission lines
from hydrogen-like, helium-like, and near-neutral ions of elements
from neon through iron \citep{nag94,ebi96}.  For
$\xi\sim10^3$\,\ergs{}cm, recombination and subsequent electronic
cascades (recombination radiation) are an efficient source of line
emission.

In several studies, X-ray emission line spectra observed with \asca\
were used to characterize the winds in HMXBs using line emissivities
due to recombination.  Two of these systems, Vela~X-1 and Cen~X-3,
were found to have very different wind characteristics that may serve
to illuminate important physical processes which occur in X-ray binary
winds.  The characteristics of the wind in Vela~X-1 were explored by
\citet{sak99} by fitting the emission-line spectrum observed during
eclipse to the recombination radiation calculated from model winds.
It was found that the emission lines from hydrogen and helium-like
lines could be fit by a wind model with a density (or equivalently,
for a given wind velocity, a mass-loss rate) approximately an order of
magnitude smaller than that inferred in other other studies using
different methods \citep[e.g.,][using the electron scattered continuum
as discussed above, and the 6.4\,keV iron fluorescence line]{lew92}.
The emission line spectrum also exhibited fluorescence lines from
near-neutral ions.  All of the observational data could be explained
if the wind had a population of dense clumps filling only a small part
of the wind volume but containing most ($\sim$90\%) of the mass of the
wind.  It is conceivable that a wind could be driven by radiation
pressure on the low-ionization material in the clumps.  By contrast,
Cen~X-3 has an X-ray luminosity of $10^{38}$\,\ergs, more than an
order of magnitude greater than that of Vela~X-1.  Its spectrum, as
seen by \asca, does not contain any fluorescence lines produced in the
wind\footnote{A fluorescent iron line feature observed from Cen~X-3 is
produced near the neutron star \citep{nag92,day93b,ebi96}}, indicating
the absence of a low-ionization component.  \citet{woj01} showed that
its emission line spectrum can be explained by recombination from a
smooth, highly ionized wind.  It was speculated that the large X-ray
flux in the wind of Cen~X-3 might evaporate clumps or prohibit them
from forming.  Radiation driving of such a smooth, highly ionized wind
does not appear plausible, and it would appear that an alternate
mechanism, such as X-ray heating of the companion is necessary
\citep[e.g.,][]{day93a}.  The winds of HMXBs have also been studied
using X-ray emission lines and the assumption that the emission lines
are due to recombination radiation by \citet{ebi96} and \citet{bor01}.

For a given X-ray luminosity and a given system geometry, the
luminosity of recombination lines from the wind of a HMXB depends on
the wind parameters only through its density distribution, as the
temperature and ionization state are determined by the ionization
parameter.  As can be seen from Equation~\ref{eqn:n}, the density, and
therefore the recombination line emissivity, depends on the mass-loss
rate and the velocity only in the combination $\mdot/v$.
\citet{woj01} has described explicitly how, for a velocity profile
which scales with the parameter $\vinf$ (the terminal velocity), the
luminosity of the recombination lines depends on $\mdot$ and $\vinf$
only in the combination $\mdot/\vinf$, and derived for Cen~X-3 a value
of $\mdot/\vinf\approx10^{-6}\,M_\sun$yr$^{-1}(1000\,{\rm
km\,s^{-1}})^{-1}$.  The degeneracy in these parameters could be
removed if velocities could be measured independently, such as through
Doppler line shifts.  The resolution of the SIS detectors was
approximately 2\% in the iron K region (6.4--7.0\,keV), and worse at
the lower energies of the other emission lines, and so was
insufficient to detect Doppler shifts of less than $\sim$2000\,\kms.
As this is of the same order as the typical terminal velocity of UV
driven winds, this has not allowed for strong constraints on the wind
driving mechanism.  The gratings on \chandra\ however, make possible
detection of Doppler shifts as small as a few hundred \kms\ and
therefore allow, in principle, tighter constraints on the wind driving
mechanism.  The HMXB Cen~X-3 is the most luminous persistent HMXB in
the galaxy and, presumably, is an extreme example of disruption of the
wind by X-rays.  It is, therefore, a strong candidate for alternative
wind driving mechanisms.  We observed Cen~X-3 over an eclipse with the
\chandra\ HETGS in order to constrain the wind velocity by resolving
emission line Doppler shifts and thus constrain the mass-loss rate and
possible wind driving mechanisms.

The spectra we obtain from our observations exhibit emission lines
from hydrogen- and helium-like ions from magnesium to iron, as well as
a a fluorescent line from near-neutral iron.  We resolve the $n=2\to1$
triplet of helium-like silicon, partially resolve the helium-like
triplet of iron, and derive constraints on the respective flux ratios
of the triplet components.  Outside of eclipse, the constraints on the
ratios are consistent with recombination radiation.  However, in
eclipse, in contrast to our expectations, the constraints are {\em
not\/} consistent with those expected for recombination.  The $w$
(resonance) lines are more intense, relative to the other components
of the triplets, than expected for recombination and are, in fact,
consistent with emission from a collisionally ionized plasma.  Similar
enhanced $w$ lines have also been seen from Vela~X-1 \citep{sch02}.

This, at first, appears to be a significant challenge to our paradigm
of HMXB winds.  Hydrogen- and helium-like line emission in
collisionally ionized plasmas occurs at much higher temperatures
($\sim$1\,keV) than in photoionized plasmas ($\sim$100\,eV) and a
mechanical source of heating, such as shocks in the wind, would be
required to maintain such a high temperature.  However, while contributions to
the line spectrum from collisionally-ionized gas cannot be ruled out a
priori, it is more natural to attribute the enhanced fluxes of lines
with large oscillator strengths to direct photoexcitation by radiation
from the neutron star.  Indeed, the dangers of neglecting
photoexcitation have long ago been pointed out for the case of optical
spectra of planetary nebulae \citep{sea68}, in which a hot star
photoionizes a cloud of surrounding gas.  More recently, clear
evidence for photoexcitation-driven line emission has also been
obtained from Seyfert 2 spectra, where similarly enhanced He-like $w$
lines have been observed with \chandra\ HETGS  \citep{sak00b}.  The
presence of absorption lines in Seyfert 1 galaxies (where a direct
line of sight to the compact radiation source exists) gives further
evidence for photoexcitation in these objects \citep{kaa00,kas01}.
X-ray spectra of Seyfert galaxies are similar to those of HMXBs,
implying that plasma in the two types of systems has similar
ionization conditions ($\log\xi\sim$2--4).  In both types of systems,
plasma column densities are of order $10^{21}$\,cm$^{-2}$.
Furthermore, in both types of systems, the X-ray emitting gas is
thought to have bulk motions with velocities of order 100 to
1000\,\kms.  Therefore, we should not be surprised that
photoexcitation is an important process in producing the observed
spectra in HMXBs.

As resonant X-ray line scattering has received little attention in the
literature, in this paper we focus on the effects of resonant
scattering in HMXBs, with the expectation that this work will
complement parallel studies in the Seyfert galaxy domain.  We show
that resonant scattering of radiation from the compact source in an
HMXB results in increased line fluxes for lines with large oscillator
strengths (such as the $w$ lines of helium-like ions) in observations
during X-ray eclipse.  We further show that this line flux enhancement
due to resonant scattering saturates as line optical depths become
comparable to unity and that the relative line fluxes that we observe
require non-zero line optical depths.  

The line optical depths along a path depend on the velocity
distribution along that path.  \citet{kin02} have calculated the
saturation of the line flux enhancement due to resonant scattering in
plasmas without bulk motions but with a Gaussian
distribution of ion velocities and shown that the \xmm\ RGS emission
line spectrum of the Seyfert 2 galaxy NGC~1068 can be fit for
appropriate values of the ion column density.  In a gas where bulk
motions are larger than the thermal or other small-scale velocities,
the optical depth, and therefore, the depletion of the resonantly
scattered line luminosity, depends on the bulk velocity distribution.
In the context of an HMXB wind, therefore, the resonantly scattered
line luminosity has a dependence on $\vinf$ other than in the
combination $\mdot/\vinf$.  We show that in spectra where the
helium-like $n=2\to1$ triplets can be resolved, the portion of line
luminosity due to resonant scattering may be derived, and, for a given
wind model, explicit constraints on the terminal wind velocity may be
derived.  These constraints may be derived only from observed line
fluxes, and do not require any information regarding Doppler line
shifts or broadening.

In \S\ref{sec:obs}, we describe our observations and reduction of the
data.  In \S\ref{sec:spec_anal_lines}, we fit the line fluxes, widths,
and shifts and derive constraints on the ratios of the fluxes of the
components of the helium-like triplets.  In \S\ref{sec:res_scat} we
calculate line emissivities due to resonant scattering, show that the
observed line ratios can be explained by resonant scattering, and
describe how constraints on the wind velocity can be derived.
In \S\ref{sec:discussion} we summarize our conclusions and discuss the
errors inherent in the previous studies which have not included the
effects of resonant scattering.

\section{Observation and Data Reduction}
\label{sec:obs}

Cen~X-3 was observed with the {\it Chandra\/} observatory on 5 March
2000 from 08:10:14 to 19:54:18 UT with the HETGS (Canizares et al.\
2002 in prep.) in place and with the ACIS-S detector array in standard
configuration.  The observation occurred over 42,444 seconds during
which time data was collected over 39,174 seconds.  Using the
ephemeris of \citet{nag92}, this is the phase interval $-$0.19 to
0.03.  

\begin{figure}
\includegraphics[angle=0,width=6.0in]{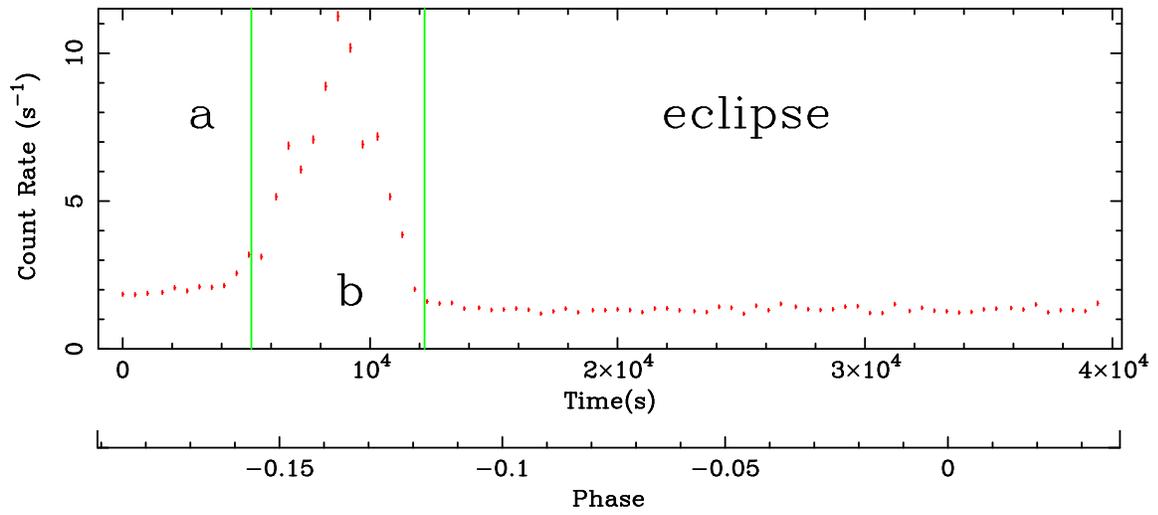}
\caption{The light curve of Cen X-3 derived from the
X-rays dispersed by the HETG.  Background has not been subtracted.
Labels ``a'', ``b'', \& ``eclipse'' indicate time segments which have
been used in the analysis. }
\label{gr_lc}
\end{figure}

\begin{figure}
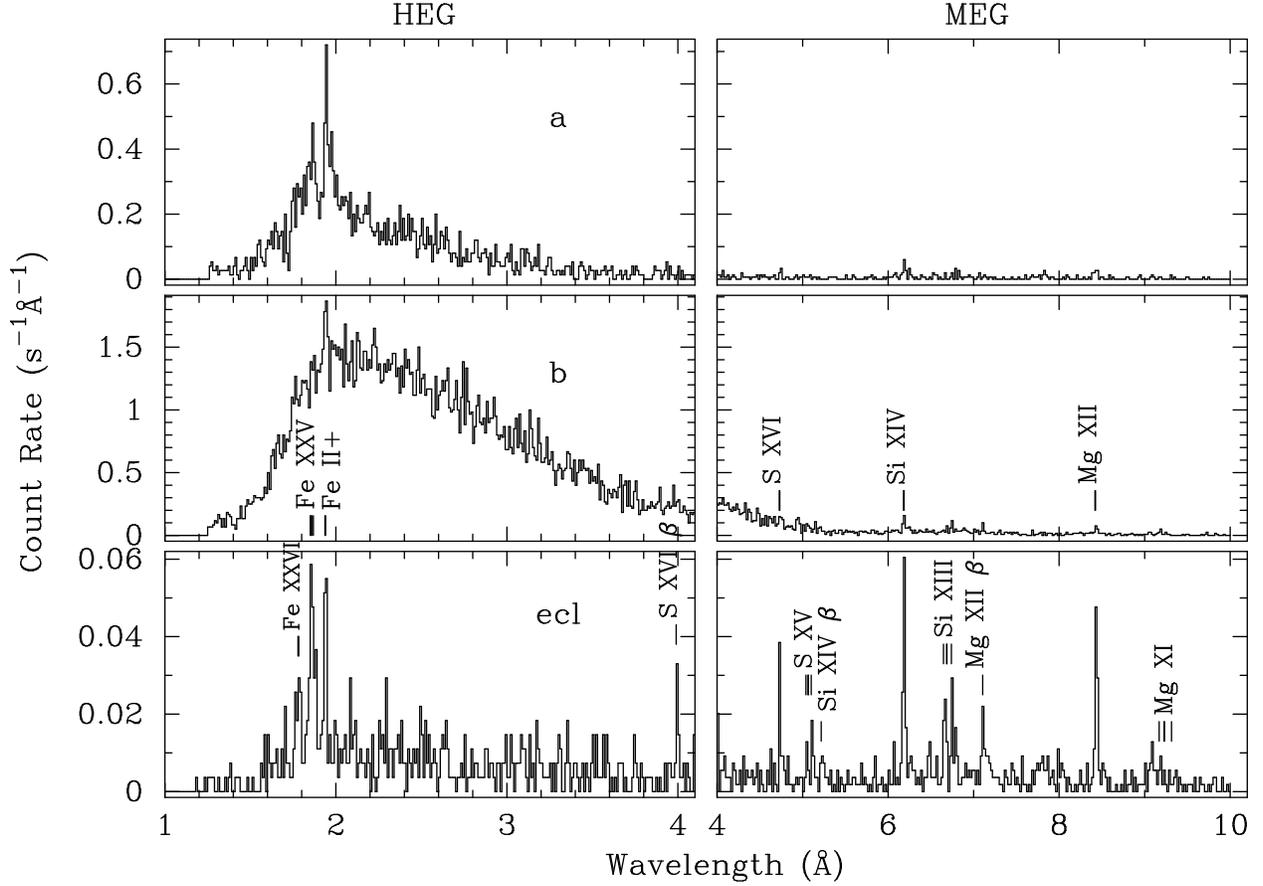

\ig{6.5}{f2.ps} 
\caption{Spectra of Cen~X-3 for the three time
segments indicated in Figure~\ref{gr_lc}.  We show the HEG spectra for
the 1--4\,\AA\ region and the MEG spectra for the 4--10\,\AA\ region.
Prominent emission lines are labels.  The labels indicate the
$n=2\to1$ transitions except for the labels containting ``$\beta$''
which indicate $n=3\to1$ transitsion.}
\label{all_spec}
\end{figure}

The pipeline data processing identified the zero-order image of
Cen~X-3 in the ACIS-S image and identified detector events due to
dispersion in the first three orders by the HETG.  From this event
list we created a light curve which we show as Figure~\ref{gr_lc}.
The flux increased, possibly as the source came out of a pre-eclipse
dip, before going into eclipse.  We divided the observation into three
time segments which we indicate on the lightcurve: ``a'' during the
dim time at the beginning of the observation (phase $-$0.19 to
$-$0.157), ``b'' during the bright time before eclipse (phase $-$0.157
to $-$0.118) and the reminder of the observation ``eclipse'' (phase
$-$0.118 to 0.03).  We extracted dispersed spectra from the entire
processed event list and also from each of the time segments indicated
above.  The spectra for orders greater than one show very few counts
and therefore we use only the first order spectra in our analysis.
We plot selected regions of the spectra from each of the time segments
in Figure~\ref{all_spec}.  The wavelength scale of the grating response
matrices used in this analysis have been found to be inaccurate by 540
parts per million due to a contraction of the detector pixel scale.
As a correction was not available for this at the time that the data
was reduced, we corrected for this simply by adjusting all of the
wavelengths we used by this amount.  When we plot spectra however, the
uncorrected scale is used.

Due to the brightness of the source, the zero-order image of Cen~X-3
suffers from pile-up and it is used in our analysis only in the
pipeline processing to find the origin of the dispersion angle
coordinate.  Pileup may affect the automatic zero-order position
determination.  However, an inspection of the detector image shows the
automatically determined position to be very close to the center of
the zeroth order image as well as the known position of the source
\citep{cla78}.

\section{Spectral Analysis}
\label{sec:spec_anal_lines}

Inspections of the first order MEG and HEG spectra show almost no
counts longward of 10\,\AA.  Lyman $\alpha$ emission lines from
hydrogen-like neon, magnesium, silicon, and sulfur can be seen.
Higher order Lyman lines and helium-like $n=2\to1$ emission lines are
also apparent from some of these elements.  An iron K$\alpha$
fluorescence emission feature is also apparent.  

We fit indiviudual Lyman $\alpha$ lines and individual helium-like
$n=2\to1$ emission complexes for elements other than iron using only
the spectral channels within 0.25\,\AA\ of those features.  Because
the iron K features are closely spaced, we fit that entire region
simultaneously using the spectral channels in the range
1.53--2.20\,\AA.  All of our spectral fits were done with the XSPEC
\citep[v11,][]{arn96} program.  In our fits, we used power laws to
model the continua and Gaussians (XSPEC model ``zgauss'') for each
emission line component.  The three parameters of the Gaussian
emission components are the flux ($I$), width (Gaussain $\sigma$), and
shift relative to the rest wavelength ($v_z$).  We express both
$\sigma$ and $v_z$ as Doppler velocities with positive values of $v_z$
corresponding to shifts to longer wavelengths (redshifts).  The
velocity of the center of mass of the Cen~X-3/V779~Cen system is only
39$\pm$4\,\kms\ \citep{hut79} and we do not corrected for this in our
analysis.  Our spectra have few counts per bin and so we fit by
minimizing the C-statistic, as implemented in
XSPEC\footnote{\url{http://xspec.gsfc.nasa.gov/docs/xanadu/xspec/manual/manual.html}},
rather than $\chi^2$, as $\chi^2$ is not an accurate goodness-of-fit
parameter for this case.  All of the errors and limits in this work
are derived using the condition $\Delta C=1$ which corresponds to the
68.3\% confidence interval.

The \lya\ line is actually a doublet.  The decay of the excited level
$2P_{3/2}$ produces the \lyaone\ line which has wavelength slightly
shorter than the \lyatwo\ line produced by the excited level
$2P_{1/2}$.  For recombination, as well as emission from thin
collisionally ionized plasmas, these two components are emitted in
proportion to their statistical weights, i.e., the luminosity of
\lyaone\ is twice that of \lyatwo.  These doublets are not resolved in
our data so it is not possible to fit them independently.  However, in
order to obtain line shifts and widths as accurately as possible, we
include both components in our fits but constrain $\sigma$ and $v_z$
for the two components to be equal and constrain $I_{\rm
Ly\alpha_1}/I_{\rm Ly\alpha_2}$ to be 2.  In Table~\ref{tab:h}, we
give the best fit parameter values and errors for our fits to \lya\
doublets other than iron using both grating sets.  We have used the
line wavelengths of \citet{joh85} as the rest-frame values.  For many
lines, the signal-to-noise ratio is poor, which may result in spurious
results (e.g., when the fitting algorithm identifies random
fluctuations of the spectrum as lines).  Therefore, we only include
lines for which the best fit line flux is at least three times the
$\Delta C=1$ lower error.

In order of increasing wavelength, the components of the helium-like
triplet are the $w$ line (also referred to as the resonance line, and
produced by the decay of the excited level $1s2p\,^1P_1$ to ground),
the $x$ and $y$ lines ($1s2p\,^3P_{2,1}\to$ground, also referred to,
together, as the intercombination line), and the $z$ line
($1s2s\,^3S_1\to$ground, also referred to as the forbidden line).  As
with the hydrogen-like \lya\ doublets, the $x$ and $y$ lines cannot be
resolved with {\it Chandra} (Hence the designation of this line
complex as a triplet) but we included both components separately and
tied the relative fluxes to those expected from recombination ---
$I_y/I_x$=10.0 for \ion{Mg}{11}, 4.0 for \ion{Si}{13}, 2.0 for
\ion{S}{15}, and 0.76 for \ion{Fe}{25}.  The relative
emissivities of the components of the triplets depend, as will be
described later, on several physical variables, so, other than the
constraint on $I_y/I_x$, we set no constraints on the flux ratios of
these lines.  We do, however, constrain all three lines to have the
same shifts and widths.  For the wavelengths of the components of the
He-like triplet, we use the line wavelengths of \cite{dra88}.

\begin{figure}
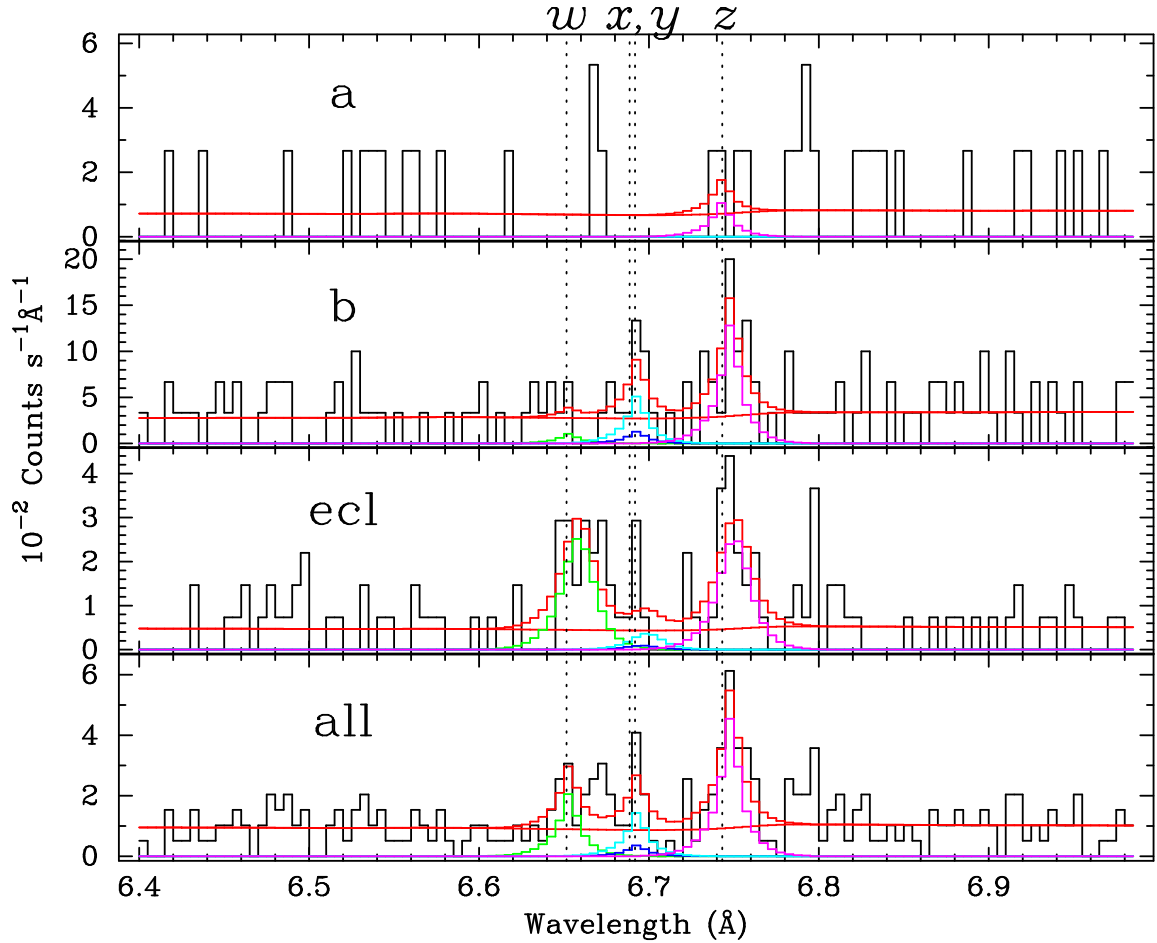

\ig{6.0}{f3.ps}
\caption{The helium-like 2$\to$1 triplet of silicon at
each of the three phases with the MEG (data in black) and the best fit
Gaussian line components. The $w$ line is plotted in green, $x$ in
blue, $y$ in light blue, and $z$ in magenta.  The rest wavelengths
(corrected for the erroneous wavelength scale, \S\ref{sec:obs}) are
indicated with dashed lines.  The power-law continuum is the red
dashed line and the total model spectra is in red.  The widths and
shifts of the three line components are constrained to vary together
but the fluxes vary independently except that the ratio of the
intercombination $y$ flux to the intercombination $x$ flux is fixed at
the values expected from recombination (4.0).  During eclipse, the
resonance line is strong.}
\label{fig:si_he}
\end{figure}

The signal-to-noise ratio of the data are not large enough to make
useful constraints on the components of the helium-like triplets of
\ion{Mg}{11} or \ion{S}{15}.  The spectral data for the \ion{Si}{13}
triplet and the best-fit model lines are plotted in
Figure~\ref{fig:si_he}.  The line fluxes, shifts, widths, and flux
ratios of these lines are shown in Table~\ref{tab:he}.  Also in this
table, we include the expected line flux ratios from recombination
radiation in a photoionized plasma as well as the expected ratio for a
collisionally ionized (coronal) plasma.  For recombination radiation,
excitation rates as a function of temperature have been derived using
HULLAC \citep{kla77} and PIC \citep{sal88}.  For the collisional case,
we use emissivities from \citet{mew85}.  In both cases, we use the
values at the temperature of peak emissivity (from \citealp{sak99},
for recombination in a photoionized plasma) though in both cases,
these ratios are only weak functions of temperature.  The line
emissivities expected for resonance scattering are discussed in
\S\ref{sec:res_scat}.

The radiative $n=2\to1$ transitions following ionization of a K
electron from an iron ion with a filled L shell may, given the
resolution of the {\it Chandra} HEG, be grouped into two line groups
analogous to the \lya\ doublet in hydrogen-like ion.  Like the \lya\
doublet, these two transition groups are separated by approximately
15\,eV (corresponding to $\approx$5\,m\AA) and have emissivities with
the ratio 2:1.  The energies (wavelengths) of the fluorescent line
groups of neighboring charge states of \ion{Fe}{2}--\ion{Fe}{17} are
shifted by approximately 15\,eV (5\,m\AA) from each other
\citep{kaa93}.  While the separations between the fluorescent line
groups for a single charge state and the separations between the line
groups of neighboring charge states are too small to be resolved by
\chandra, fluorescence emission from several different charge states
might make for an observably broadened line.  We fit the 1.94\,\AA\
emission feature using a doublet in which both components have the
same Doppler velocity and width, the flux ratio is fixed at 2:1, and
the rest wavelengths are those of \citet{bea67} for \ion{Fe}{2}
(neutral before K shell ionization).  For these fits, for all except
the eclipse phase, we additionally fixed the Doppler velocities of the
hydrogen- and helium-like lines to be zero, as the quality of the data
was not sufficient to produce meaningful fits without these
constraints.  The data and best fit models for the iron K region are
shown in Figure~\ref{fig:fe}.  The best-fit parameters, parameter
limits, and, for the helium-like triplet of iron, the expected line component ratios are tabulated in Table~\ref{tab:fe}.

The best-fit Doppler shifts and widths of lines of elements other than
iron are almost all less than 500\,\kms\, and for all measured lines,
with the single exception of \ion{Mg}{12}~\lya, consistent (within the
$\Delta C=1$ limit) with values of 500\,\kms\ or less.  For iron, the
velocities are sometimes larger, though in most of these cases, the
best fit line intensity is less than three times the $\Delta C=1$
lower error.  Therefore, the best-fit values of the width and velocity
may be spurious.  The two exceptions are the fluorescence line fit to
time segment ``a'' and the \lya\ line fit to the summed ``all''
spectrum.  In Figure~\ref{fig:fe}, it can be seen that the ``a''
spectrum contains a narrow peak near the wavelength of the
\ion{Fe}{2}~K$\alpha$ line and a smaller narrow peak approximately
40\,m\AA\ longward of the \ion{Fe}{2}~K$\alpha$ line.  Our model fails
to fit either peak individually and instead the $\sigma$ parameter for
\ion{Fe}{2}~K$\alpha$ becomes large so as to fit both peaks.  If
another line component is added to the model, both peaks can be fit as
narrow lines with the shift and width of the stronger line consistent
with less than 500\,\kms.  It may be that this peak longward of
\ion{Fe}{2}~K$\alpha$ is due to Compton recoil (Paerels, et al.\ 2002,
in prep.).  The fit to the \lya\ line ($\lambda$=1.78\,\AA) may be
affected by the absorption edge of neutral iron (1.74\,\AA) and by the
K$\beta$ neutral fluorescence line (1.76\,\AA).  We therefore conclude
that our spectrum is consistent with all lines having widths and
shifts less than 500\,\kms.

\begin{figure}
\includegraphics[angle=0,width=3.0in]{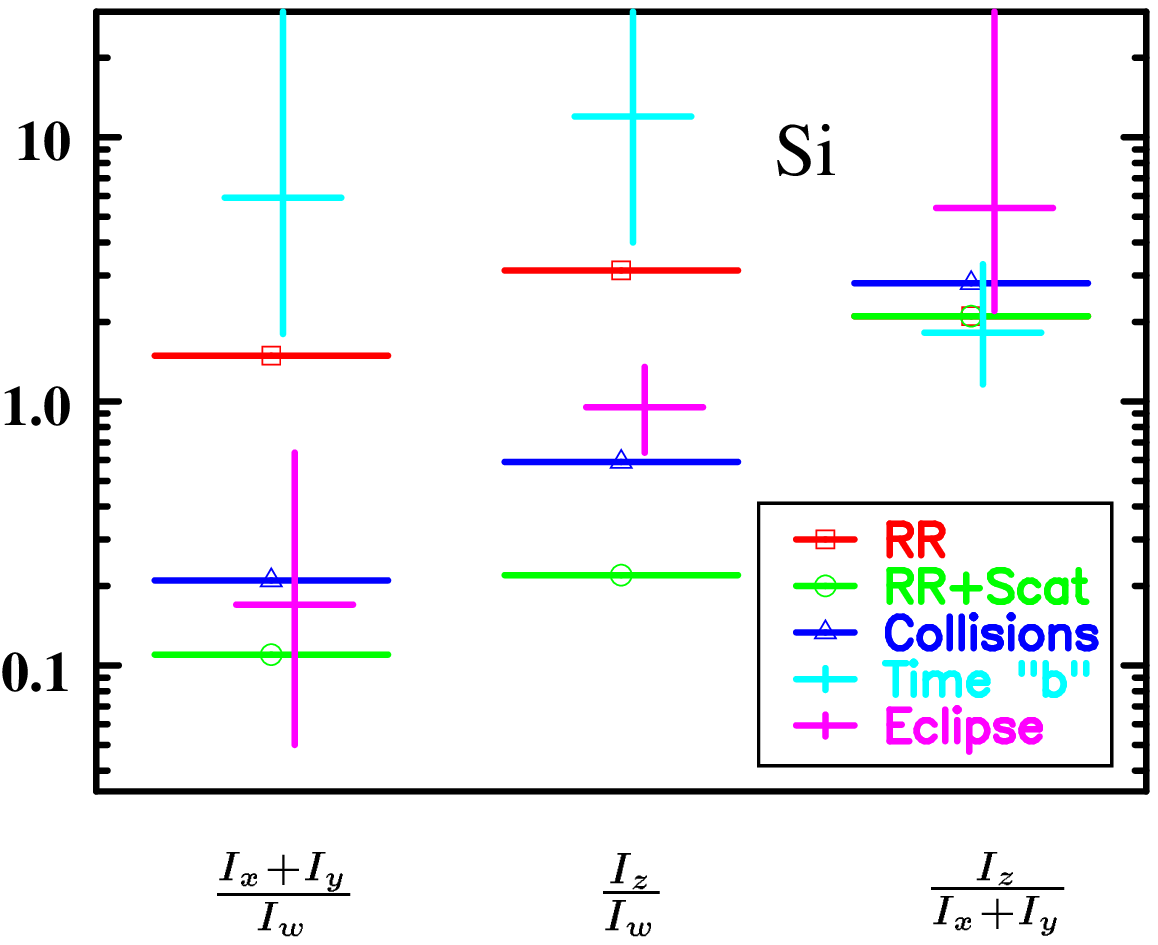}
\includegraphics[angle=0,width=3.0in]{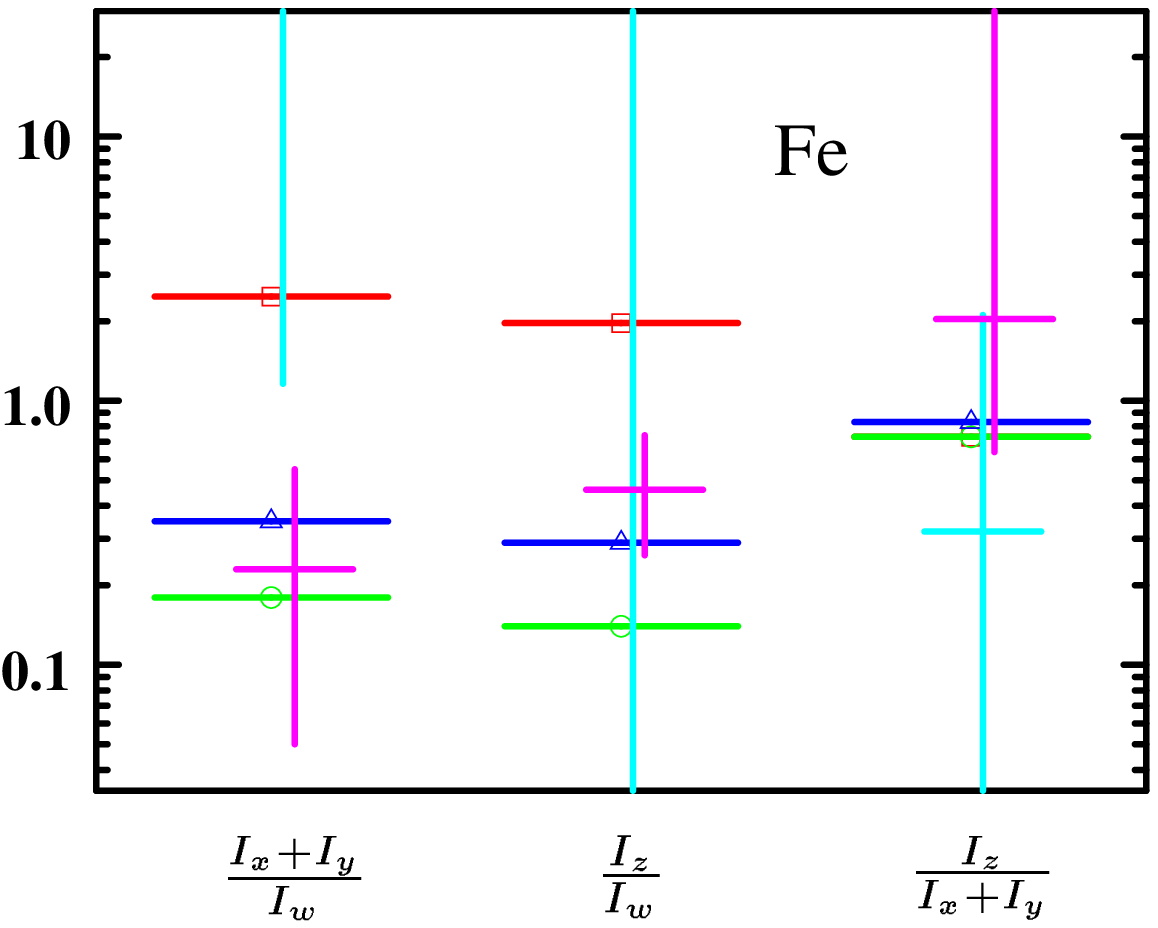} 
\caption{Emission line flux ratios for the
$n=2\to1$ complex of helium-like silicon and iron.  The expected ratios
for recombination radiation in a photoionized plasma are given in red
and marked with squares.  The expected ratios for recombination
radiation plus resonant scattering in a photoionized plasma are given
in green and marked with circles.  The expected ratios for emission
from collisional gas are given in blue and marked with triangles.  The
observed ratios (with errors) are in light blue for time interval ``b''
(before eclipse) and in magenta for eclipse.}
\label{fig:rat}
\end{figure}

We show the limits derived for the line component flux ratios of the
helium-like triplets of silicon and iron in and out of eclipse, as
well as the expected ratios for emission from a plasma in collisional
equilibrium in Figure~\ref{fig:rat}.  It would appear from the
observed line ratios that the wind seen during eclipse is
collisionally ionized and that the wind seen outside of eclipse is
photoionized.  This would be puzzling, owing to the fact that during
the two orbital phases, much of the same wind is visible to the
observer. However, until now we have not considered resonant
scattering.

\section{Resonant Scattering}
\label{sec:res_scat}

\begin{figure}
\includegraphics[width=2.6in]{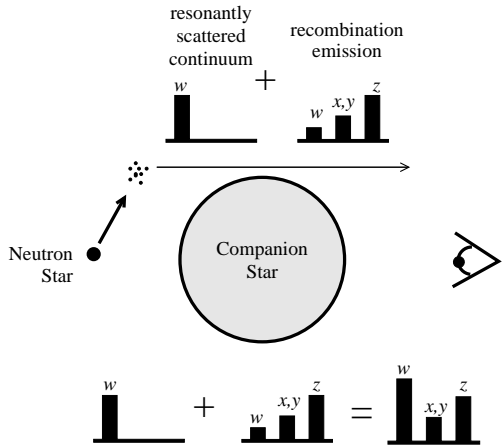}
\caption{In eclipse, only X-rays which have been reprocessed in the
visible part of the wind are observed.  Observable line photons are
produced by recombination and also resonant scattering.  In the
helium-like triplets, the resulting $w$ line is stronger, relative to
the other lines of the triplet, than would be expected from pure
recombination.} 
\label{fig:ecl_scat}
\end{figure}

\begin{figure}
\includegraphics[width=2.6in]{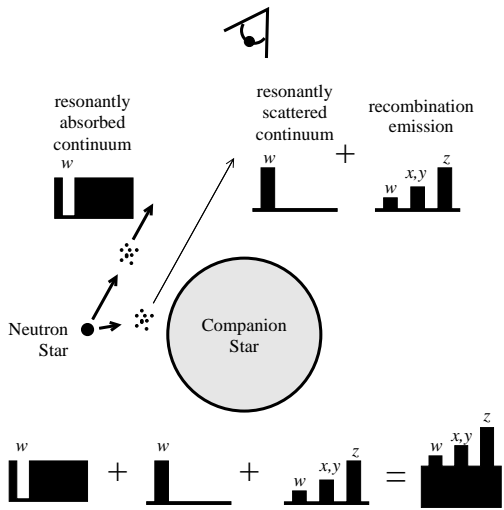}
\caption{Outside of eclipse, the observer sees line emission due to
recombination and resonant scattering in the bulk of the wind as in
eclipse.  However, the observer also sees the continuum directly from
the neutron star which has a ``notch'' due to resonant scattering
along the line of sight.  In the sum of the observed radiation, the
notched continuum tends to offset the line emission due to resonant
scattering in the bulk of the wind.}
\label{fig:unecl_scat}
\end{figure}

Transitions which have significant large oscillator strengths can be
excited by direct photoexcitation.  This additional rate of excitation
leads to an increase in the rate of line emission as the photoexcited
ions decay.  When one electron from a ground state hydrogen or
helium-like ion is photoexcited to the $n=2$ level, the decay is
almost always the inverse process of the photoexcitation.  We refer to
this photoexcitation and subsequent decay as resonant scattering.  The
\lya\ lines of hydrogen-like ions and the $n=2\to1$ $w$ lines of
helium-like atoms have large oscillator strengths but the $x$, $y$,
and $z$ lines of helium-like atoms do not.  In the wind of an X-ray
binary, photoexcitation by radiation from the compact source will
therefore result in enhanced line emission of lines with large
oscillator strengths.  In eclipse, when the wind is visible, but the
neutron star is not, resonant scattering results in an enhancement of
the $w$ line, but not the $x$, $y$, and $z$ lines, of the helium-like
triplet.  Out of eclipse, however, when the neutron star is visible,
photons near the rest frequency from the neutron star will be
scattered out of the line of sight to the observer, resulting in a
continuum with a ``notch'' from the line of sight to the neutron star.
If the velocities in the material along the line of sight to the
neutron star are not too different from those in the bulk of the gas,
the emission line will be superimposed on the notch, reducing the
apparent line flux.  Resonant scattering, then, represents a plausible
mechanism for increasing the flux of the $w$ line during eclipse but
not outside of eclipse.  We illustrate this schematically in
Figures~\ref{fig:ecl_scat} and \ref{fig:unecl_scat}.  

In order to determine whether resonant scattering can quantitatively
explain the effects we observe, we calculate the line luminosity
enhancement due to resonant scattering in a distribution of gas
surrounding a point source of radiation.  In hydrogen and helium-like
ions, electrons may also be excited to the $n=2$ state by
photoexcitation to $n>2$ states and subsequent decay.  However, as
will be described later, this process represents a small contribution
to the excitation rate and we will neglect it here.  Since resonant
scattering, as defined here, does not create or destroy photons, The
resonantly scattered line luminosity is equal to the luminosity
removed from the primary radiation:
\begin{equation}
L_{\rm scat}=L_{\nu_0}(4\pi)^{-1}\int\!\!\!\int (1-e^{-\tau_{\rm px}})d\nu
d\Omega .
\label{eqn:lscat}
\end{equation}
where $L_{\nu_0}$ is the specific luminosity of the point source at
$\nu_0$, the line rest frequency.  We have assumed here that $L_\nu$,
the specific luminosity of the point source, is not a strong function
of frequency in the neighborhood of the line rest frequency.  The
line (photoexcitation) optical depth from the compact radiation source
to infinity, $\tau_{\rm px}$, is a function of frequency and direction
and depends on the line oscillator strength and the density and
velocity distributions of the scattering ion.  In principle, the
frequency integral runs from 0 to $\infty$ but in this work we are
considering velocities that are small compared to $c$, and the
integrand will be non-zero only in a small interval near $\nu_0$, the
line rest frequency.  

For a plasma in photoionization equilibrium, the recombination line
emissivity may be written in terms of the photoionization rate
\citep[e.g.,][]{ost89}.  It will further be useful to express the
photoionization cross-section in terms of an oscillator strength.
\citet{bet57} define the quantity $df/dE$ such that $(df/dE)\delta E$
is the summed oscillator strength for absorption transitions to
continuum levels between $E$ and $E+\delta E$.  We use $h\nu=\chi+E$,
where $\chi$ is the ionization potential, and define $df/d\nu$ which
differs from $df/dE$ by the factor $h^{-1}$.  The photo quantity
$df/d\nu$ may be derived from the quantity $df/dE$, which is defined
such that differs from the quantityis similar to may be defined as
follows:
\begin{equation}
4\pi j_{\rm rec}=h\nu_0 n_{Z,z}\eta
\frac{\pi e^2}{mc}  
\int_\nuth^\infty \frac{4\pi}{h\nu}\frac{L_\nu}{4\pi s^2}\frac{df}{d\nu}d\nu.
\label{eqn:jrec_f}
\end{equation}
Here $j_{\rm rec}$ is the recombination line emissivity, the energy
emitted in the line due to recombination per unit volume, time, and
solid angle.  The symbol $h$ is Planck's constant, $\nu_0$ is the
line rest frequency,  $e$ and $m$ are the electronic charge and mass,
and $c$ is the speed of light.  The symbols $Z$ and $z$ are, respectively, the
atomic number and charge of the ion from which the recombination line
is emitted and $n_{Z,z}$ is the density of that ion.  The quatity
$L_\nu/4\pi s^2$ is the mean intensity of radiation (energy per unit
area per unit time per unit frequency per steradian, averaged over
solid angle, assuming that the plasma is optically thin in the
photoionizing continuum) due to the point radiation source, and $\eta$ is the
fraction of recombinations that result in emission of the line.  The
value of $\eta$ depends on temperature, but only weakly.

The recombination line luminosity is then 
\begin{equation}
L_{\rm rec}=\int\!\!\!\int4\pi j_{\rm rec}s^2ds d\Omega=
h\nu_0\eta\frac{\pi e^2}{mc}
\int\!\!\!\int n_{Z,z} ds d\Omega
\int_\nuth^\infty\frac{L_\nu}{h\nu}\frac{df}{d\nu}d\nu 
\label{eqn:lrec_intn}
\end{equation}
It will be useful to relate the spatial integral to the
photoexcitation optical depth, which may be done as follows: 
\begin{equation}
f\frac{\pi e^2}{mc}\int n_{Z,z}ds=\int\tau_{\rm px}d\nu
\label{eqn:tau_col}
\end{equation}
where $f$ is the line oscillator strength.
Equation~\ref{eqn:lrec_intn} then becomes 
\begin{equation}
L_{\rm rec}=
h\nu_0\eta f^{-1}\int\!\!\!\int\tau_{\rm px}(\nu^\prime)d\nu^\prime d\Omega
\int_\nuth^\infty\frac{L_\nu}{4\pi h\nu}\frac{df}{d\nu}d\nu
\label{eqn:lrec}
\end{equation}
where we have noted the dependendence of $\tau_{\rm px}$ on
$\nu^\prime$ explicitly.

Dividing Equation~\ref{eqn:lscat} by Equation~\ref{eqn:lrec} gives
\begin{equation}
\frac{L_{\rm scat}}{L_{\rm rec}}=\frac{f L_{\nu_0}}{h\nu_0\eta
\int_\nuth^\infty(L_\nu/h\nu)(df/d\nu)d\nu}\,
\frac{\int\!\!\!\int(1-e^{-\tau_{\rm px}})d\nu^\prime
d\Omega}{\int\!\!\!\int\tau_{\rm px}d\nu^\prime d\Omega} 
.\label{eqn:scat/rec}
\end{equation}
We define the last fraction in this equation as the optical depth
functional $g(\tau_{\rm px})$.  Its value ranges from 0 to 1, and if
the column is optically thin in the line ($\tau_{\rm px}\ll1$ for all
$\nu$ and all directions), its value approaches unity.  The first
fraction on the right hand side then gives the value of $L_{\rm
scat}/L_{\rm rec}$ in the optically thin limit.

The variation of the line luminosity contributions due to
recombination and resonant scattering with optical depth may be
understood as follows.  While the line luminosity due to resonant
scattering increases with the line optical depth, it saturates as the
line optical depth becomes comparable to unity.  Recombination also
saturates, but with the optical depth of the photoionization
continuum.  As the cross-section in the photoionization continuum is
generally much smaller (so small that we have neglected it) than the
line cross-section, resonant scattering is always more saturated than
recombination).  Therefore, the relative contribution due to resonant
scattering is greatest when neither process is saturated, i.e., when
the line optical depth is small.

If the neutron star is not occulted, we observe a notched continuum
from along the line of sight to it, as mentioned above.  We add a term
to Equation~\ref{eqn:lscat} to account for the resulting reduced
apparent luminosity.
\begin{equation}
L_{\rm scat,app}= L_{\nu_0}\int\left[(4\pi)^{-1}\int (1-e^{-\tau_{\rm px}})
d\Omega -(1-e^{-\tau_{\rm px}({\mathbf \Omega}_{\rm obs})})\right] d\nu 
\end{equation}
where ${\mathbf \Omega}_{\rm obs}$ indicates the line of sight to the
observer.  Adding the absorption term to Equation~\ref{eqn:scat/rec}
is straightforward.  It can be seen that for spherical symmetry, the
apparent value of $L_{\rm scat, app}$ is zero, as expected --- i.e.,
resonant scattering makes no net contribution to the observed line
fluxes.

In a real HMXB, the companion star may occult a significant fraction
of the wind and so the relations derived above are not accurate.  It
is straightforward to generalize the above expressions for line
luminosities for the case of an occulting body of finite size.
However, this extra mathematical complexity does not serve to
illustrate the points we wish to make here.  The expression for
$L_{\rm scat}/L_{\rm rec}$, discounting the gas behind the star, is
still the optically thin value derived above multiplied by a quantity
ranging from zero to one, which depends on optical depths and is one
in the optically thin limit.  Therefore, we continue with the
approximation that the entire wind is visible to the observer.

\subsection{Resonant Scattering Effects in Hydrogen and Helium-like
Ions}

The oscillator strengths of the $x$, $y$, and $z$ lines are very small
and so it is a good approximation to neglect any contribution to the
emissivity those lines due to resonant scattering.  For the
hydrogen-like \lya\ doublet and the helium-like $w$ line
$f\approx0.4(Z-z)$.  The value of ($Z-z$) is equal to the number of
electrons in the ion.  The sum of the oscillator strengths of all of
the transitions from ground to levels with $n>2$ is only approximately
0.1$(Z-z)$, justifying, for our approximate treatment here, our
neglect of these transitions.  The variation of $df/d\nu$ with $\nu$
is described approximately by $df/d\nu=\kappa \nu^{-3}$ where, for the
ground state of hydrogen and helium-like ions, the normalization
constant $\kappa$ is given by
\begin{equation}
\int_\nuth^\infty\frac{df}{d\nu}d\nu\approx0.5(Z-z).
\end{equation}
Evaluating this integral gives $\kappa\approx(Z-z)\nuth^2$.  If
$L_\nu\propto\nu^{-\gamma}$
, then
\begin{equation}
\int_\nuth^\infty \frac{L_{\nu}}{h\nu}\frac{df}{d\nu}d\nu\approx
\frac{L_{\nu_0}}{h\nuth}(Z-z)\left(\frac{\nu_0}{\nuth}\right)^\gamma(\gamma+3)^{-1}
\end{equation}
and Equation~\ref{eqn:scat/rec} becomes:
\begin{equation}
\frac{L_{\rm scat}}{L_{\rm rec}}\approx
0.4(\gamma+3)\left(\frac{\nuth}{\nu_0}\right)^{\gamma+1}\eta_{\rm
line}^{-1} \,g(\tau_{\rm px}) 
\label{eqn:scat/rec_numa}
\end{equation}
For Cen~X-3, we take $\gamma=0$ \citep[approximately the value
determined by ][]{san98}\footnote{The quantity $F_\nu$ is the energy
flux, so $\gamma$ is related to the exponent $\alpha$ (index of the
photon flux power-law) by $\gamma=\alpha-1$.}.  Using this and
$(\nuth/\nu_0)\approx4/3$, we have, for the hydrogen $n=2\to1$ line
and the helium-like $n=2\to1$ resonance line:
\begin{equation}
\frac{L_{\rm scat}}{L_{\rm rec}}\approx
1.6\eta^{-1} \,g(\tau_{\rm px}).
\label{eqn:scat/rec_numb}
\end{equation}  

Photoexcitations to bound states above the $n=2$ level and resulting
cascades also contribute to the luminosity of the $n=2\to1$ lines.
However, the oscillator strength summed over all levels $n>2$ is only
approximately $0.1(Z-z)$.  Therefore, in the optically thin limit, the
photoexcitation rate to all states with $n>2$ is much less than the
rate to the $n=2$ state.  As the $n=2\to1$ transition becomes
saturated, the photoexcitation rate to levels with $n>2$ may approach
that of the $n=2$ level.  However, the spontaneous deexcitation rates
for photoexcited $n>2$ levels to excited levels are generally less
than the deexcitation rates to the ground state by approximately a
factor of ten.  Therefore, even if the $n=2\to1$ line becomes highly
saturated, the contribution to luminosities of the $n=2\to1$ lines
from photoexcitations to the $n=2$ level will dominate over
contributions to photoexcitations to $n>2$ levels.  Therefore, for the
purposes of this paper, we neglect such photoexcitations to $n>2$
levels.  Equations~\ref{eqn:scat/rec_numa} and~\ref{eqn:scat/rec_numb}
may be derived with more precise atomic data.  However, due to fact
that that our measurements of the line flux ratios have large
statistical errors, and the fact that that we have not considered the
effects of cascades following photoexciation to levels with $n>2$,
this is not likely to result in a qualitative improvement in our
results. 

Using HULLAC \citep[Hebrew University and Lawrence Livermore Atomic
Code, ][]{kla77} as described in previous work
\citep{lie96,sak99,woj01} we have, for ions with $Z$=6--26, found the
value of $\eta$ for the $w$ line of helium-like ions to be 0.12 and
the value of $\eta$ for the \lya{} doublets to be 0.41.  For our
functional form of $F_\nu$, in the optically thin limit, resonant
scattering results in a contribution to the $w$ line luminosity
approximately 13 times that of recombination.  However, resonant
scattering makes only a negligible contribution to the $x$, $y$, and
$z$ lines, so that resonant scattering results in a contribution to
the total $n=2\to1$ line luminosity approximately 2.3 times that of
recombination.  We show in Tables~\ref{tab:he} and \ref{tab:fe} and
plot in Figure~\ref{fig:rat}, the expected line ratios for
recombination plus resonant scattering in the optically thin limit
along with the expected values for emission from a collisionally
ionized plasma and with the measured values.  While the relative
fluxes of the $w$ lines in eclipse are larger than expected for
radiative recombination, the $z/w$ ratios indicate that the relative
fluxes of the $w$ lines are less than expected for recombination
radiation plus resonant scattering in the optically thin limit.  If
the line emission is due to recombination and resonant scattering,
non-zero optical depths of the $w$ lines are indicated.  For the
\lya{} doublets, resonant scattering in the optically thin limit
provides a line emissivity approximately 3.9 times that of
recombination.  However, from this unresolved doublet alone, it is not
possible to measure what fraction of the observed line flux is due to
resonant scattering nor is it possible to estimate the implied optical
depth.

\subsection{Observational Constraints on Line Optical Depth}

\label{sec:obs_const}
The luminosity of the $w$ line due to resonant scattering depends on
an integral of the line optical depth over solid angle and frequency.
It is not possible, from the measured line flux ratios, to extract the
complete optical depth function.  However, if the optical depth, is
zero for all frequencies and directions, except for some finite region
in frequency and directions where it has the single positive value
$\tau_0$, the $g$ functional simplifies to $g=(1-e^{-\tau_0})/\tau_0$.
Such a distribution of optical depth would be realized, for example,
by a cone with its vertex at the radiation source with constant ion
density inside the cone and a constant velocity gradient away from the
radiation source.  While we do not expect the wind to have this simple
distribution of optical depth, the value of $\tau_0$ for which the
observed line ratios are reproduced represents a characteristic
optical depth for the wind.

\begin{figure}
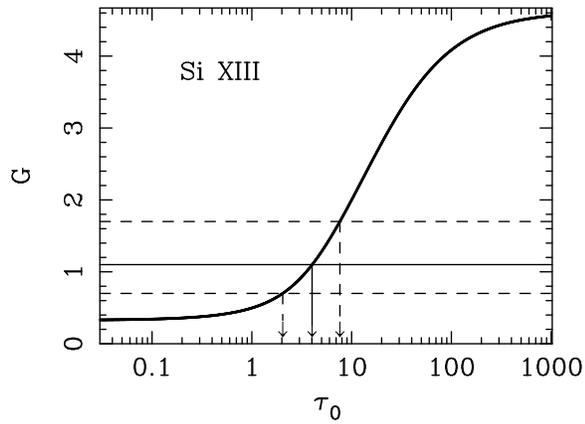

\ig{3.0}{f8.ps}
\caption{The $G$ ratio (thick line) for the \ion{Si}{13} $n=2\to1$
triplet including resonant scattering for a gas distribution in which
the optical depth from the primary radiation source to infinity is
$\tau_0$ in within a range of frequency and solid angle and zero
otherwise.  Our measured value of this $G$ ratio in eclipse ---
G=1.1\ud{0.6}{0.4}, indicated by the horizontal lines implies a value
for this characteristic optical depth.}
\label{fig:g}
\end{figure}

Because resonant scattering enhances the luminosity of the $w$ line,
but not the luminosities of the $x$, $y$, or $z$ lines, it is
convenient to examine the effects of resonant scattering using the
quantity
\begin{equation}
G\equiv\frac{I_x+I_y+I_z}{I_w}
\end{equation}
In Figure~\ref{fig:g}, we plot the value of the $G$ ratio for
\ion{Si}{13} as a function of $\tau_0$ along with our measured value
of $G$ and the implied value of $\tau_0$.  Because the value of $G$
for pure recombination ($G_{\rm rec}$) is known and the $x$, $y$, and
$z$ lines are not affected by resonant scattering, it is possible to
determine how much of the observed flux in the $w$ line is due to
recombination radiation and how much is due to resonant scattering.
In Table~\ref{tab:G}, for \ion{Si}{13} and \ion{Fe}{25}, we give
$G_{\rm rec}$, the expected enhancement of the $w$ line due to
resonant scattering in the optically thin limit, the expected value of
$G$ for recombination plus resonant scattering, and the following
quantities determined for the eclipse phase: $G$, $I_{w,{\rm scat}}$,
the ratios of $I_{w,{\rm scat}}$ to $I_{w,{\rm rec}}$ and to the total
$n=2\to1$ flux due to recombination, and $\tau_0$.

\section{Discussion}

\label{sec:discussion}

We have observed Cen~X-3 from before an eclipse until mid-eclipse.
During this observation, the continuum flux underwent a large
increase, then decreased due to the eclipse ingress.  We have fit the
shifts, widths, and fluxes of observable lines.  Our best fit velocity
shifts and (Gaussian $\sigma$) velocity widths are generally less than
500\,\kms.  These velocities are significantly smaller than terminal
velocities of isolated O star winds \citep[1--2$\ee3$\,\kms,
e.g.,][]{lam99}.  However, in isolated O stars, X-ray line velocity
widths (HWHM or Gaussian $\sigma$) have been observed from
$\sim$400\,\kms \citep{sch00} to $\sim1000$\,\kms\
\citep{wal01,cas01,kah01a}.  Our results for the Doppler velocities
are therefore consistent with those in isolated O stars.

We have measured the ratios of the fluxes of the components of the
helium-like triplets of silicon and iron.  We find that the flux
ratios are consistent with recombination in a photoionized plasma,
with the exception that during the eclipse phase, the $w$ line is
stronger than expected from recombination.  Resonant scattering
provides a natural mechanism for increasing the fluxes of the $w$
lines in eclipse but not out of it.  We have calculated the
enhancement of the $w$ line fluxes due to resonant scattering and we
find that the observed relative $w$ line fluxes are smaller than
expected for resonant scattering in the optically thin limit and
non-zero line optical depths are required.  We are therefore able to
explain the observed emission line triplets of helium-like ions ---
which, during eclipse, appear very much like what would be observed
from a hot, collisionally ionized plasma --- without rejecting the
hypothesis that the wind is composed entirely of a warm
photoionionized plasma. 

As mentioned in \S\ref{sec:intro}, in previous spectral studies of
HMXB winds with moderate-resolution spectral data, X-ray emission line
spectra have been interpreted with the assumption that line emission
is due purely to recombination radiation.  The primary physical
quantity inferred from a line luminosity is, as the emissivity due to
recombination depends on the square of the density, an emission
measure: $\int n_e^2dV$.  If the apparent line emissivity is
underestimated, such as by neglect of resonant scattering, then the
inferred emission measure is overestimated by that same amount.  The
parameter $\mdot/\vinf$ is proportional to the wind density and
therefore proportional to the square root of the wind emission
measure.  We have shown that outside of eclipse, resonant scattering
along the line of sight to the neutron star offsets the line flux
enhancement due to resonant scattering in the bulk of the gas.  In our
case, however, these two effects appear to have may nearly cancelled,
and so for interpreting spectra obtained outside of eclipse, it may be
a good approximation to ignore resonant scattering.  For spectra
obtained during eclipse, however, resonant scattering, in the
optically thin limit, increases the $n=2\to1$ line fluxes of hydrogen-
and helium-like ions by factors of, respectively 3.9 and 2.3.
Therefore, emission measures inferred assuming line emissivities from
recombination only, from spectra obtained during eclipse, could be too
large by factors as large as four.  However, in fact, the error due to
neglect of resonant scattering is probably not so large as the winds
may have significant optical depths in the lines, which, as discussed
earlier, decreases the line luminosities due to resonant scattering
relative to the line luminosities due to recombination.  Indeed, as we
have shown in \S\ref{sec:res_scat}, for the eclipse spectrum described
here, the total $n=2\to1$ flux due to resonant scattering for
\ion{Si}{13} and \ion{Fe}{25} is only approximately 0.6 and 1.0,
respectively that due to recombination.  Therefore, previous estimates
of the parameter $\mdot/\vinf$ may be too large only by factors of
about two or less.  While this does not qualitatively
the affect conclusions reached in any previous studies, the effects of
resonant scattering should be included, even in the analysis of
spectra of medium-resolution spectra, if only so the basic wind
parameters may be derived accurately.

While the neglect of resonant scattering in previous work may not have
resulted in significant errors in derived wind parameters,
interpretation of high-resolution X-ray spectra with proper
consideration of resonant scattering allows the determination of wind
parameters that cannot otherwise be determined.  As \citet{woj01} have
shown, for a typical model wind, the wind density, and therefore the
luminosity of recombination line emission depends on the parameters
$\mdot$ (the mass-loss rate) and $\vinf$ (the terminal wind velocity)
only in the combination $\mdot/\vinf$ and that it is therefore
impossible to constrain either of these parameters individually using
only the observed luminosities of lines resulting from recombination.
Resonant scattering, however, has a different dependence on these
parameters.  Emission from resonant scattering results as photons from
the compact object are redirected by scattering ions toward the
observer.  We imagine a model wind for which we decrease the velocity
parameter $\vinf$ while keeping the value of $\mdot/\vinf$, and
therefore the density, constant.  As $\vinf$ is decreased, the Doppler
shifts decrease and the same number of ions scatter photons from a
decreasing frequency range.  This results in a decrease in the number
of photons which are subject to scattering but an increase in the
scattering optical depth and an increase in the fraction of the
accessible photons which can be scattered.  However, the fraction of
the accessible photons which may be scattered is, of course, limited
to be no more than unity and as $\vinf$ is decreased, the total
resonantly scattered line luminosity decreases.  Because $\mdot/\vinf$
is kept constant, however, the line luminosity due to recombination
remains constant.  The emission line luminosities due to recombination
and resonant scattering, therefore, depend on $\mdot$ and $\vinf$ in a
non-degenerate way.  If the line luminosities due to recombination
and resonant scattering can be determined, independent constraints on
$\mdot$ and $\vinf$ can be derived.  We have demonstrated that, with
high resolution spectroscopy, it is possible to resolve helium-like
triplets and thereby discriminite between the line emission due to
recombination and resonant scattering.  Therefore, using high
resolution X-ray spectroscopic data, it is possible to derive
independent constraints on the parameters $\mdot$ and $\vinf$.
However, deriving these constraints requires detailed modeling of the
wind, which is beyond the scope of this work, so we do not derive such
constraints here.  However, this effect is most sensitive in the
regime where the line optical depth is of order unity.  As we have
shown in \S~\ref{sec:obs_const}, the line optical depths of the $w$
lines of helium-like silicon and iron in Cen~X-3 are, in fact, of
order unity and so the prospects for using this effect to derive
independent constraints on the wind parameters $\mdot$ and $\vinf$ are
good.  Furthermore, these constraints do not depend at all on
resolving Doppler line shifts or broadenings and are therefore
independent of any constraints which may be obtained from observed
line profiles.

From our analysis of \chandra\ spectra of Cen~X-3, we may conclude
that in the analysis of high-resolution spectroscopic data, such as is
obtained using the gratings on \chandra\ and \xmm, of the winds of
X-ray binaries and X-ray photoionized plasmas in general, it is
critical that the effects of resonant scattering be included, if
observed spectra are to be reproduced, or if inferences are to be made
regarding the conditions of the emitting plasma (e.g., for determining
whether plasma is photoionized, collisionally ionized, or ionized by a
``hybrid'' of both processes).  Furthermore, in the analysis of
high-resolution spectroscopic data from HMXBs, including the effects
of resonant scattering not only allows wind parameters to be derived
accurately, but allows new constraints on wind parameters.

\acknowledgements

We thank Ali Kinkhabwala for useful discussions regarding resonance
scattering.  We thank David Cohen for many useful comments on the
manuscript.  This research has made use of NASA's Astrophysics Data
System Abstract Service.  D.\ A.\ L.\ was supported in part by NASA
Long Term Space Astrophysics Grant S-92654-F.  MS was partially
supported by NASA through Chandra Postdoctoral Fellowship Award Number
PF1-20016 issued by the Chandra X-ray Observatory Center, which is
operated by the Smithsonian Astrophysical Observatory for and behalf
of NASA under contract NAS8-39073.  Work at LLNL was performed under
the auspices of the U.S. Department of Energy, National Nuclear
Security Administration by the University of California, Lawrence
Livermore National Laboratory under contract No. W-7405-Eng-48.

\bibliographystyle{apj} 
\bibliography{ms}

\begin{deluxetable}{lcccccccccc}
\tabletypesize{\small}
\tablewidth{0pt}
\tablecaption{Fit Parameters from \lya\ Doublet Fits}
\tablehead{ Time & Grating & Species & $\sigma$(\kms) &
$v_z$(\kms) 
& $I$\tablenotemark{a}}
\startdata
a & MEG  & Mg & 800\ud{400}{200} &0$\pm${300} &1.3\ud{0.5}{0.4} \\
& & Si & $<${200} & 80\ud{250}{\phn20} &1.4\ud{0.5}{0.4} \\
b & MEG & Mg &400\ud{300}{200} &300$\pm${200} &2.1\ud{0.7}{0.6} \\
& & Si &100\ud{200}{100} &270\ud{130}{220} &3.5\ud{0.9}{0.8} \\
ecl & MEG  & Mg &210$\pm${100} &340$\pm${70} &1.3$\pm${0.2} \\
& & Si &300\ud{200}{300} &160$\pm${100} &1.6\ud{0.3}{0.2} \\
& & S &$<${300} &0\ud{290}{\phn30} &2.2\ud{0.6}{0.4} \\
& HEG  & Mg &120\ud{140}{120} &290$\pm${80} &1.3\ud{0.4}{0.3} \\
& & Si &420\ud{140}{120} &320$\pm${130} &1.7\ud{0.4}{0.3} \\
& & S &$<${200} &280\ud{\phn30}{260} &2.2\ud{0.6}{0.5} \\
all & MEG  & Mg &340\ud{130}{110} &300\ud{80}{90} &1.5$\pm${0.2} \\
& & Si &10\ud{130}{\phn10} &280\ud{\phn20}{220} &2.0\ud{0.3}{0.2} \\
& & S &300\ud{400}{300} &230\ud{280}{220} &2.9\ud{0.7}{0.6} \\
& HEG  & Mg &$<${300} &320$\pm${80} &1.0\ud{0.3}{0.2} \\
& & Si &$<${200} &530\ud{\phn30}{100} &1.5$\pm${0.3} \\
& & S &$<${700} &310\ud{\phn40}{120} &2.6\ud{0.7}{0.6} \\
\enddata
\tablenotetext{a}{Line intensity units are $10^{-5}$\,ph\,cm$^{-2}$\,s$^{-1}$}
\label{tab:h}
\end{deluxetable}

\begin{deluxetable}{lllcccccccccc}
\tabletypesize{\scriptsize}
\tablewidth{0pt}
\tablecaption{He-like Si $n=2\to1$ MEG data}
\tablehead{ time
& $v_z$(\kms) & $\sigma$(\kms) &
$I_w$\tablenotemark{a} & $I_x+I_y$\tablenotemark{a} &
$I_z$\tablenotemark{a} & 
 $(I_x+I_y)/I_w$ & $I_z/I_w$ & $I_z/(I_x+I_y)$ 
}
\startdata
a & 0(fixed) & 0(fixed) & 0.0\ud{0.1}{0.0} & 0.0\ud{0.1}{0.0} &
0.2$\pm${0.2} & \nodata & $>$0.86 & $>$1.0 \\
b & 120\ud{70}{40} & 0\ud{40}{\phn0} & 0.2\ud{0.4}{0.2} & 1.0\ud{0.6}{0.4}
& 1.8\ud{0.7}{0.5} & 6.2\ud{\infty}{4.8} & 11.5\ud{\infty}{8.5} &
1.8\ud{2.1}{0.8} \\
ecl & 300\ud{140}{160} & 240\ud{130}{230} & 0.5$\pm${0.2} &
0.1$\pm${0.1} & 0.5$\pm${0.1} & 0.2\ud{0.4}{0.2} &
0.9\ud{0.5}{0.3} & 5.0\ud{\infty}{3.3} \\
all & 100\ud{20}{10} & 0\ud{30}{\phn0} & 0.3$\pm${0.1} & 0.3$\pm${0.1} &
0.6\ud{0.2}{0.1} & 0.9\ud{0.7}{0.5} & 2.1\ud{1.7}{0.8} &
2.3\ud{1.9}{0.8} \\
\hline
process(es) \\
\hline
rec(75\,eV) & & & & & &  1.49 & 3.13 & 2.1 \\
rec+scat(75\,eV) & & & & & &  0.11 & 0.22 & 2.1 \\
col(900\,eV) & & & & & & 0.21 & 0.59 & 2.8 \\
\enddata
\tablenotetext{a}{Line intensity units are $10^{-5}$\,ph\,cm$^{-2}$\,s$^{-1}$}
\label{tab:he}
\end{deluxetable}

\begin{deluxetable}{lllcccccccccc}
\tabletypesize{\scriptsize}
\tablecaption{Iron line data}
\tablewidth{0pt}
\tablehead{
 &\multicolumn{4}{c}{} & \multicolumn{3}{c}{processes} \\
\cline{6-8}
 &\multicolumn{4}{c}{time} & rec & rec+scat &
collisions \\
\cline{2-5}
parameter &a &b &ecl &all &  (250\,eV) & (250\,eV) & (5\,keV) \\ } 
\startdata
\cutinhead{\lya}
 $v_z$(\kms) & 0(fixed) &0(fixed) &-640\ud{800}{780} &0(fixed)  \\
 $\sigma$(\kms) &3700\ud{1300}{1000} &4600\ud{1900}{1300} &1600\ud{800}{600} &4000\ud{1100}{\phn900}  \\
 $I$ &35\ud{13}{12} &70$\pm$30 &3.7\ud{1.3}{1.2} &20\ud{6}{5}  \\
\cutinhead{He-like}
 $v_z$ &0(fixed) &0(fixed) &160\ud{20}{40} &0(fixed)  \\
 width &2900\ud{\phn800}{1000} &200\ud{1100}{\phn200} &$<${60}
&2100\ud{1200}{1500} \\ 
 $I_w$ &20\ud{40}{20} &$<${4.9} &4.3\ud{1.7}{1.4} &$<${15.3}  \\
 $I_x+I_y$ &40\ud{30}{40} &23\ud{14}{11} &0.9\ud{1.4}{0.9} &17\ud{\phn9}{17}
\\ 
 $I_z$ &$<${29} &12\ud{16}{12} &2.1\ud{1.2}{0.8} &4\ud{10}{\phn4}
\\ 
 $(I_x+I_y)/I_w$ &\nodata &$>$0.86 &0.2\ud{0.7}{0.2} &\nodata  & 2.48 & 0.18 & 0.35 \\
 $I_z/I_w$ &\nodata &\nodata &0.5\ud{0.4}{0.2} &\nodata  & 1.97 & 0.14 &
0.29 \\
 $I_z/(I_x+I_y)$ &\nodata &0.5\ud{3.2}{0.5} &2.2\ud{\infty}{1.6} &\nodata
& 0.73 & 0.73 & 0.83 \\
\cutinhead{K$\alpha$}
 $v_z$ &2000\ud{600}{500} &100$\pm$400 &80\ud{200}{\phn60} &350$\pm${200}  \\
 width &3200$\pm$600 &1100\ud{500}{400} &20\ud{330}{\phn20} &900$\pm$300 \\
 $I$ &80$\pm$11 &76\ud{17}{16} &4.0\ud{1.0}{0.9} &24$\pm$3 \\
\enddata
\tablecomments{Units of $v_z$ and $\sigma$ are \kms.  Units of $I$
are 10$^{-5}$\,ph\,cm$^{-2}$s$^{-1}$}
\label{tab:fe}
\end{deluxetable}

\begin{deluxetable}{lcc}
\tablewidth{0pt}
\tablecaption{Helium-like triplet quantities}
\tablehead{ quantity 
		& \ion{Si}{13} 		& \ion{Fe}{25} }
\startdata
$G_{\rm rec}$ 	& 4.62			& 4.45 \\
$\left(\frac{I_{w,{\rm scat}}}{I_{w,{\rm rec}}}\right)_{\tau=0}$
		& 13			& 13 \\
$G_{\rm rec+scat,\tau=0}$ 
		& 0.33			& 0.32 \\
$G$		& 1.1\ud{0.6}{0.4}	& 0.7\ud{0.8}{0.3} \\
$I_{w,{\rm scat}}$
		& 0.4$\pm$0.2		& 3.6\ud{1.7}{1.4} \\
$\frac{I_{w,{\rm scat}}}{I_{w,{\rm rec}}+I_x+I_y+I_z}$
		& 0.6\ud{0.4}{0.3}	& 1.0\ud{0.9}{0.6} \\
$\frac{I_{w,{\rm scat}}}{I_{w,{\rm rec}}}$
		& 3.2\ud{2.4}{1.5}	& 5.3\ud{4.9}{3.2} \\
$\tau_0$	& 4.0\ud{3.6}{2.0}	& 2.2\ud{4.0}{1.7} \\
\enddata
\tablecomments{Values without errors are theoretical values.  Values
with errors are derived from the eclipse phase.}
\label{tab:G}
\end{deluxetable}

\clearpage

\clearpage

\clearpage

\clearpage

\begin{figure}
\ig{6.0}{f4.ps}
\caption{The iron K region with the HEG. The hydrogen-like
\lyaone\ and \lyatwo\ lines at 1.78\,\AA\ are in green and blue,
respectively, as are the K$\alpha_1$ and K$\alpha_2$ fluorescence
lines at 1.94\,\AA.  The \lyatwo\ and K$\alpha_2$ lines are
constrained to have the same widths and half the flux of,
respectively, the \lyaone\ and K$\alpha_1$ lines.  The legend and
constraints on the the He-like $2\to1$ lines are otherwise same as
those in Figure~\ref{fig:si_he} except that the $y$ to $x$ flux
ratio for iron is 0.76.  Again the resonance line is strong during
eclipse. }
\label{fig:fe}
\end{figure}

\clearpage

\clearpage

\clearpage

\end{document}